# Output Feedback Control of Linear Time-Invariant Systems with Operational Constraints

Marcel Menner[1], Heather Hussain[2], Eugene Lavretsky[3]
*The Boeing Company, USA*

**This paper introduces a systematic method for designing robust linear controllers using output feedback in the presence of operational constraints. The design uses Nagumo's Theorem and the Comparison Lemma to guarantee constraint satisfaction, while incorporating min-norm optimal control principles inspired by Control Barrier Functions. The resulting controller is a continuous piecewise-linear output feedback policy that preserves the closed-loop system's analyzability using linear systems theory. Due to the linear control design, multi-input multi-output (MIMO) robustness margins can be derived with and without active operational constraints. This paper shows that operational constraints on the system's state can be satisfied using an observer-based output feedback control design. Through flight control trade studies, we demonstrate the practical relevance of the framework in safety-critical aircraft control applications.**

## I. Introduction

Robustness and control margins are critical aspects to ensure operational safety and practical functionality [1]. The foundation of these analyses is linear system theory [2], [3], which assumes that linear models sufficiently capture the behavior of the dynamical system and its control system. Although linear approximations are generally valid within typical operating envelopes, their accuracy can deteriorate when actuators saturate or when outputs exceed prescribed limits. To handle such nonlinearities, heuristic strategies like anti-windup schemes are frequently implemented to improve adverse transient effects during saturation [4].

This work extends our prior contributions [5], [6] by introducing a structured method for designing output feedback controllers for multi-input multi-output (MIMO) linear time-invariant (LTI) systems subject to operational constraints. We focus on observer-based output feedback controllers that enforce component-wise box constraints. The design leverages key theoretical tools: the Nagumo Theorem [7] (English translation [8]), the Comparison Lemma [9], and principles from min-norm optimal control [10]. The Nagumo Theorem provides conditions for forward invariance of the closed-loop trajectories, while the Comparison Lemma helps formulate constraints on the limited output that the LTI system can satisfy. These constraints are incorporated into a Quadratic Program (QP), enabling systematic enforcement of operational limits through output feedback. Importantly, the QP with a specific choice of cost function weights admits a closed-form analytical solution [11] resulting in a continuous piecewise-linear output feedback control law. This explicit form facilitates rigorous analysis of system stability and robustness margins using linear system theory. We derive robustness margins for MIMO systems under active constraints within the output feedback context. Our results highlight effective strategies that guarantee bounded output tracking errors and maintain closed-loop stability despite constraints. The practical utility of the approach is demonstrated through flight control trade study.

Related methodologies include Control Barrier Functions (CBFs) [12], [13], [14] and model predictive control (MPC) [15], both of which have been extensively studied for enhancing safety in autonomous systems. For example, MPC is often used as a supervisory "safety filter" that allows a nominal controller to operate freely until constraints are approached [16]. In contrast, our approach integrates vector-valued CBFs within a min-norm optimal output feedback control framework to derive an explicit analytical control law. The resulting piecewise-linear output feedback controller can be thoroughly analyzed using classical control theory. While sharing theoretical foundations with CBF methods—namely the Nagumo Theorem, the Comparison Lemma, and min-norm optimal control—our approach applies min-norm optimal control only during the design phase, avoiding online optimization during operation. To recognize the origins and contributions of CBF theory, we refer to our method as the "CBF output feedback augmentation."

---

[1] Boeing Designated Expert – Autonomy, Aurora Flight Sciences (A Boeing Company).
[2] Associate Technical Fellow, The Boeing Company.
[3] Senior Principal Technical Fellow, The Boeing Company. AIAA Fellow.

## II. Problem Formulation

Consider the stabilizable MIMO LTI dynamical system,

$$\begin{aligned} \dot{x}(t) &= Ax(t) + Bu(t) \\ y(t) &= Cx(t) + Du(t) \\ y_{\lim}(t) &= C_{\lim}\, x(t) \end{aligned} \tag{1}$$

where $x(t) \in R^n$ is the state vector, $u(t) \in R^m$ is the control inputs, $y(t) \in R^{n_y}$ is the vector of measured outputs, and $y_{\lim}(t) \in R^m$ is the vector of limited outputs. This paper proposed an output feedback controller in the presence of operational constraints on the selected limited outputs $y_{\lim}(t)$,

$$\begin{gathered} y_{\lim}^{\min} \le y_{\lim}(t) \le y_{\lim}^{\max} \\ \Updownarrow \\ h_{\min}(x(t)) = y_{\lim}^{\min} - C_{\lim}\, x(t) \le 0 \\ h_{\max}(x(t)) = C_{\lim}\, x(t) - y_{\lim}^{\max} \le 0 \end{gathered} \tag{2}$$

with the component-wise minimum and maximum limits (box constraints), $y_{\lim}^{\min}$ and $y_{\lim}^{\max}$, respectively. Note that $C$ and $C_{\lim}$ do not have to be identical, i.e., the method in this paper can limit a state that is not directly measurable.

Consider the control command,

$$u(t) = u_{\mathrm{bl}}(t) + \pi(t) \tag{3}$$

with a control augmentation signal $\pi(t)$ and the baseline controller $u_{\mathrm{bl}}(t)$. The main goal of the method in this paper is to derive the control augmentation signal $\pi(t)$ that becomes active only if needed to avoid violating the operational constraints in (2).

***Definition* 1 (CBF-ability).** Let $r_i$ be the relative degree of the $i^{\text{th}}$ limited output, i.e., $(y_{\lim})_i$ in (1) has to be differentiated $r_i$ times until the CBF augmentation $\pi(t)$ appears explicitly, see Section III-A for details. Further, let

$$H_x = \begin{bmatrix} (C_{\lim})_1 \prod_{j=1}^{r_1} (A - \lambda_{1j} I_n) \\ \vdots \\ (C_{\lim})_m \prod_{j=1}^{r_m} (A - \lambda_{mj} I_n) \end{bmatrix}, \quad H_\pi = \begin{bmatrix} (C_{\lim})_1 A^{r_1 - 1} \\ \vdots \\ (C_{\lim})_m A^{r_m - 1} \end{bmatrix} B, \tag{4}$$

where $(C_{\lim})_i$ is the $i^{\text{th}}$ row of $C_{\lim}$. Then, system (1) is said to be CBF-able if $H_\pi$ is nonsingular and there exist negative real eigenvalues $\lambda_{ij} < 0$ such that $A - BH_\pi^{-1}H_x$ is Hurwitz.

We make the following assumptions.

***Assumption* 1**. The matrix pair $(A, B)$ is stabilizable, and the matrix pair $(A, C)$ is observable.

***Assumption* 2**. The dynamical system (1) has a well-defined vector relative degree, see Section III-A for details.

***Assumption* 3**. The dynamical system (1) is CBF-able.

### A. Controller Formulation and Problem Statement

Consider the control command in (3) with the baseline controller

$$u_{\mathrm{bl}}(t) = -K\hat{x}(t), \tag{5}$$

with feedback gain matrix $K$ and the state estimate $\hat{x}(t)$ provided by the observer dynamics

$$\begin{aligned} \dot{\hat{x}}(t) &= A\hat{x}(t) + Bu(t) + L\left(y(t) - \hat{y}(t)\right) \\ \hat{y}(t) &= C\hat{x}(t) + Du(t) \end{aligned} \tag{6}$$

with Luenberger gain $L$ [17], [18], [19] and output estimate $\hat{y}(t)$. Both the feedback gain matrix and the Luenberger gain can, e.g., be designed using a linear quadratic regulator (LQR) design.

This paper derives a piecewise-affine control augmentation policy of the form

$$\pi(t) = -K_{i,\mathrm{CBF}}\hat{x}(t) + F_i y_{\lim}^{\min/\max}, \tag{7}$$

with the CBF feedback gain matrix $K_{i,\text{CBF}}$ and the CBF command matrix $F_i$ switching to accommodate constraints, such that the overall control input $u(t)$ is continuous, closed-loop stability is guaranteed, and the limited output constraints (2) are satisfied forward in time. Compared to [5], this paper uses the state estimate $\hat{x}(t)$ instead of the state $x(t)$ to enforce (2). Fig. 1 illustrates the block diagram of the controller design.

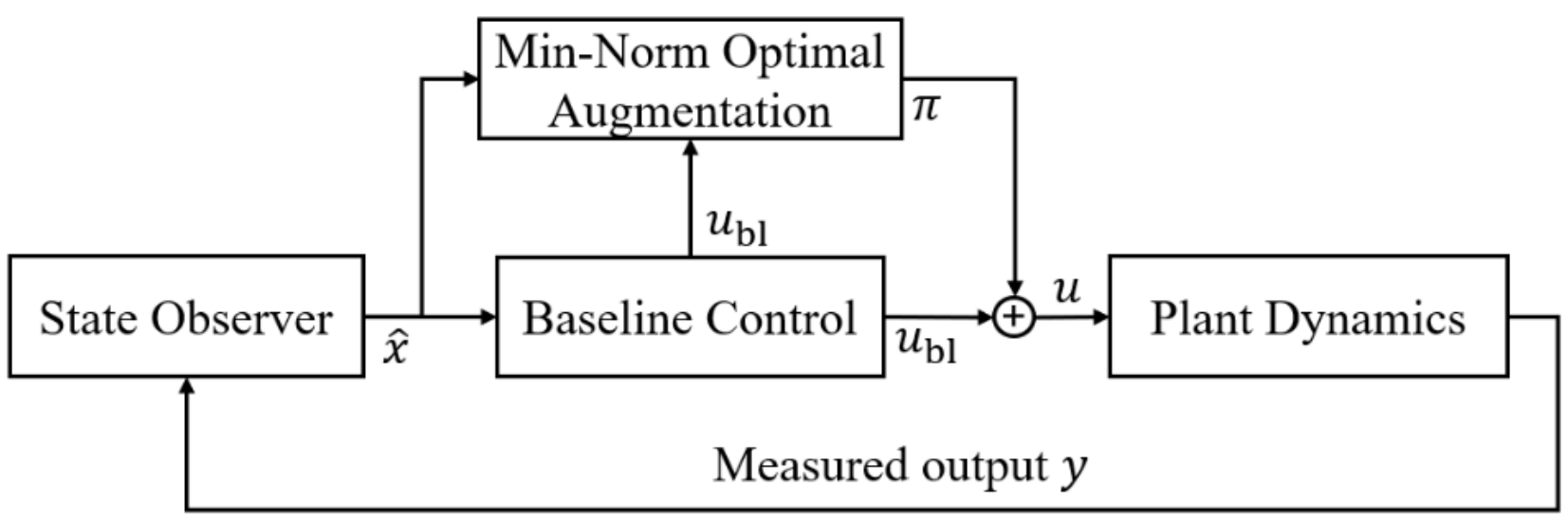

**Fig. 1. Closed-loop system block-diagram. The baseline controller is augmented such that the limited output stays within operational limits.**

### B. System under Baseline Control

We define the state estimation error

$$e_{\text{est}}(t) = x(t) - \hat{x}(t) \tag{8}$$

where, under baseline control, the closed loop system dynamics of the state and estimation error are given by

$$\begin{aligned} \dot{x}(t) &= Ax(t) - BK\hat{x}(t) = Ax(t) - BK\left(x(t) - e_{\text{est}}(t)\right) \\ \dot{e}_{\text{est}}(t) &= (A - LC)e_{\text{est}}(t) \end{aligned} \tag{9}$$

with eigenvalues given by $\lambda_{\text{obs}} = \text{eig}(A - LC)$ and $\lambda_{\text{bl}} = \text{eig}(A - BK)$, i.e., observer eigenvalues and controller eigenvalues are separated (separation principle), which can be easily seen by reformulating (9) in matrix form

$$\begin{bmatrix} \dot{x}(t) \\ \dot{e}_{\text{est}}(t) \end{bmatrix} = \begin{bmatrix} A - BK & BK \\ 0 & A - LC \end{bmatrix} \begin{bmatrix} x(t) \\ e_{\text{est}}(t) \end{bmatrix} \tag{10}$$

What remains is to design the output feedback-based control augmentation signal in (3), which is addressed next.

***Remark*** **1 (Observer-based control with loop transfer recovery (OBLTR),** [2]**).** While robustness margins generally differ between the state and output feedback case, OBLTR is a theoretical foundation that can be used to design the observer gain matrix. OBLTR uses a squaring up procedure and the algebraic Riccati equation (ARE) to compute the observer gain $L_v$ where $v > 0$ is the key design parameter defining the $Q$ and $R$ matrices in the ARE. In the limit $v \to 0$, OBLTR recovers linear quadratic regulator optimal state feedback margins at the system input break point.

## III. Control Augmentation Design

The main challenge in satisfying the operational constraints (2) is that the control input does not appear explicitly, i.e., (2) cannot be manipulated directly with the control input. To address this challenge, we leverage the Comparison Lemma [9] in order to modify the operational limits (2) and derive a control augmentation policy. Hence, the proposed method is derived to satisfy operational constraints (2) by means of manipulating modified constraints.

### A. Modified Constraints for Controller Design

Let the relative degree of the *i*-th limited output $y_{\text{lim},i}(t)$ be $r_i \geq 1$ for $i = 1, \ldots, m$, where relative degree refers to the number of time $y_{\text{lim},i}(t)$ needs to be differentiated for the control augmentation signal to appear explicitly. Suppose that $y_{\text{lim}}$ has a vector relative degree $r = (r_1 \ \ldots \ r_m)$ [20], i.e.,

$$\left[\left\|\nabla_\pi\left(y_{\text{lim},i}^{(k)}\right)\right\| = 0, \forall 1 \leq k \leq r_i - 1\right] \wedge \left[\left\|\nabla_\pi\left(y_{\text{lim},i}^{(r_i)}\right)\right\| \neq 0\right]. \tag{11}$$

Then, the control-to-limited output sensitivity matrix $H_\pi \in R^{m \times m}$ must be nonsingular, see Assumption 2,

$$H_\pi = \begin{bmatrix} (C_{\lim})_1 A^{r_1-1}B \\ \vdots \\ (C_{\lim})_m A^{r_m-1}B \end{bmatrix}, \quad \det(H_\pi) \neq 0. \tag{12}$$

For every limited output $i = 1,...,m$, the proposed control design uses the vector relative degree definition to differentiate the operational constraints $r_i$ times as follows. Consider a stable polynomial of order $r_i$, with the real roots $\{\lambda_{ij}\}_{j=1,\ldots,r_i}$ with $\lambda_{ij} < 0$,

$$\phi_i(s) = \prod_{j=1}^{r_i}(s - \lambda_{ij}) = \sum_{j=0}^{r_i} c_{ij}\, s^j, \tag{13}$$

where $c_{ij}$ denotes the $j^{\text{th}}$ coefficient of the $i^{\text{th}}$ polynomial, and by definition, $c_{i0} > 0$, for every $i = 1,\ldots,m$. The stable polynomials in (13) are used to modify the constraints as

$$\begin{gathered} Y_{\lim}(t) = \underbrace{\begin{bmatrix} \phi_1(s) & \ldots & 0 \\ \vdots & \ddots & \vdots \\ 0 & \ldots & \phi_m(s) \end{bmatrix}}_{\Phi(s)} y_{\lim} = \Phi(s)\, y_{\lim}(t) \\ H(x(t), u_{bl}(t), \pi(t)) = \begin{bmatrix} H_{\min}(x(t), u_{\text{bl}}(t), \pi(t)) \\ H_{\max}(x(t), u_{\text{bl}}(t), \pi(t)) \end{bmatrix} = \begin{bmatrix} \Phi(s) h_{\min}(x(t)) \\ \Phi(s) h_{\max}(x(t)) \end{bmatrix} = \begin{bmatrix} \Phi(s)\left(y_{\lim}^{\min} - y_{\lim}(t)\right) \\ \Phi(s)\left(y_{\lim}(t) - y_{\lim}^{\max}\right) \end{bmatrix} \leq 0, \end{gathered} \tag{14}$$

where the stable polynomials $\phi_i(s)$ are treated as differentiation operators with respect to $s$. Then, Lemma 1 in [5] shows that the modified output $Y_{\lim}(t)$ and modified constraints can be written as

$$\begin{gathered} Y_{\lim}(t) = H_x x(t) + H_\pi\left(u_{\text{bl}}(t) + \pi(t)\right) \\ \begin{bmatrix} H_{\min}(x(t), u_{\text{bl}}(t), \pi(t)) \\ H_{\max}(x(t), u_{\text{bl}}(t), \pi(t)) \end{bmatrix} = \begin{bmatrix} -Y_{\lim}(t) + \alpha_\pi\, y_{\lim}^{\min} \\ Y_{\lim}(t) - \alpha_\pi\, y_{\lim}^{\max} \end{bmatrix} = \begin{bmatrix} -H_\pi \\ H_\pi \end{bmatrix} \pi(t) + \begin{bmatrix} \Delta H_{\min}(x(t), u_{\text{bl}}(t)) \\ \Delta H_{\max}(x(t), u_{\text{bl}}(t)) \end{bmatrix} \leq 0 \end{gathered} \tag{15}$$

with

$$\begin{aligned} \Delta H_{\min}(x(t), u_{\text{bl}}(t)) &= -H_x x(t) - H_\pi u_{\text{bl}}(t) + \alpha_\pi\, y_{\lim}^{\min} \\ \Delta H_{\max}(x(t), u_{\text{bl}}(t)) &= H_x x(t) + H_\pi u_{\text{bl}}(t) - \alpha_\pi\, y_{\lim}^{\max} \end{aligned} \tag{16}$$

and

$$H_x = \begin{bmatrix} (C_{\lim})_1 \prod_{j=1}^{r_1}(A - \lambda_{1j} I_n) \\ \vdots \\ (C_{\lim})_m \prod_{j=1}^{r_m}(A - \lambda_{mj} I_n) \end{bmatrix}, \quad H_\pi = \begin{bmatrix} (C_{\lim})_1 A^{r_1-1}B \\ (C_{\lim})_2 A^{r_2-1}B \\ \vdots \\ (C_{\lim})_m A^{r_m-1}B \end{bmatrix} \tag{17}$$

are the system state and control sensitivity matrices respectively, and $\alpha_\pi \in R^{m\times m}$ is a diagonal matrix, with its positive diagonal elements defined as

$$\alpha_\pi = \begin{bmatrix} c_{10} & \ldots & 0 \\ \vdots & \ddots & \vdots \\ 0 & \ldots & c_{m0} \end{bmatrix}, \quad c_{i0} = \prod_{j=1}^{r_i}(-\lambda_{ij}) > 0, \quad \forall i = 1,\ldots,m. \tag{18}$$

Lemma 2 in [5] showed that if a dynamical system is invariant forward in time with respect to the modified constraints (15), then the dynamical system in (1) is also invariant forward in time with respect to the original constraints (2). Hence, designing the control augmentation to satisfy (15) will enforce (2) for all times.

**B. QP with Modified Operations Constraints and Control Augmentation Solution for Output Feedback**

Since the control augmentation appears explicitly in the modified constraints (15), it can directly be used for control design. Motivated by the min-norm controller design method [10], consider the following QP [11]:

$$
\begin{aligned}
&\text{Cost}: J(v,w) = \left(\pi^T R_\pi \pi\right) \to \min_\pi \\
&\text{Constraints}: H(x,u_{bl},\pi) = \begin{bmatrix} H_{\min}(x,u_{\text{bl}},\pi) \\ H_{\max}(x,u_{\text{bl}},\pi) \end{bmatrix} = \begin{bmatrix} -H_\pi \\ H_\pi \end{bmatrix}\pi + \begin{bmatrix} \Delta H_{\min}(x,u_{\text{bl}}) \\ \Delta H_{\max}(x,u_{\text{bl}}) \end{bmatrix} \le 0
\end{aligned} \tag{19}
$$

i.e., we seek the min-norm control augmentation that only becomes active (meaning $\pi \neq 0$) if needed to satisfy the modified constraints in (19). Hence, the modified constraints are constructed such than any $\pi$ satisfying (19) also makes the dynamical system forward invariant with respect to the operational constraints. Further, choosing the cost function weight $R_\pi = H_\pi^T H_\pi$, (19) has an explicit closed-form solution, which is shown in Theorem 2 in [5]. The state-dependent control augmentation policy optimizing the QP in (19) is given by

$$
\begin{aligned}
\pi(x(t)) &= H_\pi^{-1}\left(\max\left(0_{m\times 1}, \Delta H_{\min}\left(x(t), u_{\text{bl}}(t)\right)\right) - \max\left(0_{m\times 1}, \Delta H_{\max}\left(x(t), u_{\text{bl}}(t)\right)\right)\right) \\
&= H_\pi^{-1}\begin{cases} \left(-H_x x(t) - H_\pi u_{\text{bl}}(t) + \alpha_\pi y_{\text{lim}}^{\min}\right), & \text{if } \left(-H_x x(t) - H_\pi u_{\text{bl}}(t) + \alpha_\pi y_{\text{lim}}^{\min}\right) > 0 \\ -\left(H_x x(t) + H_\pi u_{\text{bl}}(t) - \alpha_\pi y_{\text{lim}}^{\max}\right), & \text{if } \left(H_x x(t) + H_\pi u_{\text{bl}}(t) - \alpha_\pi y_{\text{lim}}^{\max}\right) > 0 \\ 0, & \text{otherwise} \end{cases}
\end{aligned} \tag{20}
$$

with $H_x$ and $H_\pi$ as in (16) and $\alpha_\pi$ as in (18). For the state-feedback case, Theorem 2 in [5] further shows forward invariance with respect to both the modified constraints (15) and consequently the operational constraints, and the integrated output tracking error remains uniformly ultimately bounded (UUB) forward in time. In this paper, we extend these findings and focus on output feedback.

### C. Overall Control Design using Output Feedback

For output feedback, (20) needs to be expressed in terms of the state estimate. We use the following control augmentation policy

$$
\begin{aligned}
\pi(\hat{x}(t)) &= H_\pi^{-1}\left(\max\left(0_{m\times 1}, \Delta H_{\min}\left(\hat{x}(t), u_{\text{bl}}(t)\right)\right) - \max\left(0_{m\times 1}, \Delta H_{\max}\left(\hat{x}(t), u_{\text{bl}}(t)\right)\right)\right) \\
&= H_\pi^{-1}\begin{cases} \left(-H_x \hat{x}(t) - H_\pi u_{\text{bl}}(t) + \alpha_\pi y_{\text{lim}}^{\min}\right), & \text{if } \left(-H_x \hat{x}(t) - H_\pi u_{\text{bl}}(t) + \alpha_\pi y_{\text{lim}}^{\min}\right) > 0 \\ -\left(H_x \hat{x}(t) + H_\pi u_{\text{bl}}(t) - \alpha_\pi y_{\text{lim}}^{\max}\right), & \text{if } \left(H_x \hat{x}(t) + H_\pi u_{\text{bl}}(t) - \alpha_\pi y_{\text{lim}}^{\max}\right) > 0 \\ 0, & \text{otherwise} \end{cases}
\end{aligned} \tag{21}
$$

It is clear that $\pi(\hat{x}(t)) = \pi(x(t))$ when the state estimation error has converged, $e_{\text{est}}(t) = 0$. However, during transients with $e_{\text{est}}(t) \neq 0$, compared to the state feedback case in (20), the operational constraints are not always satisfied by construction for any choice of design parameters. Instead, for the output feedback-based control augmentation policy (21), attention needs to be paid to selecting the design parameters appropriately. Section IV shows that constraint satisfaction is achieved if the design parameters are chosen based on the observer eigenvalues.

Eq. (21) can also be expressed in the form (7). To see this, we reformulate the control augmentation solution. Let

$$
\delta(\hat{x}(t)) = \begin{bmatrix} \delta_1(\hat{x}(t)) & \dots & 0 \\ \vdots & \ddots & \vdots \\ 0 & \dots & \delta_m(\hat{x}(t)) \end{bmatrix} \in R^{m\times m} \tag{22}
$$

defining a diagonal positive-definite state-dependent matrix, whose binary-valued diagonal elements are given by the conditions in (21)

$$
\delta_i(\hat{x}(t)) = \begin{cases} 1, & \text{if } \left[\left(-H_x \hat{x}(t) - H_\pi u_{\text{bl}}(t) + \alpha_\pi\, y_{\text{lim}}^{\min}\right)_i > 0\right] \vee \left[\left(H_x \hat{x}(t) + H_\pi u_{\text{bl}}(t) - \alpha_\pi\, y_{\text{lim}}^{\max}\right)_i > 0\right] \\ 0, & \text{otherwise} \end{cases} \tag{23}
$$

$\forall i = 1,\ldots,m$. Hence, the notation $\delta_i(\hat{x}(t))$ represents a continuous switching logic based on active constraints, and, by definition, $\left\|\delta(\hat{x}(t))\right\| \le 1$ uniformly in $\hat{x}(t)$. Then, the total controller shows a modified baseline control part, a CBF feedback part, and a CBF command part:

$$
\begin{aligned}
u_{\mathrm{bl}}(t)+\pi(t) &= u_{\mathrm{bl}}(t)-H_\pi^{-1}\delta\left(\hat{x}(t)\right)\left(H_x\hat{x}(t)+H_\pi\, u_{\mathrm{bl}}(t)-\alpha_\pi y_{\mathrm{lim}}^{\mathrm{min/max}}\right) \\
&= \underbrace{\left(I_{(m)}-H_\pi^{-1}\delta\left(\hat{x}(t)\right)H_u\right)u_{\mathrm{bl}}(t)}_{\text{Modified Baseline Control}}-\underbrace{H_\pi^{-1}\delta\left(x(t)\right)H_x\hat{x}(t)}_{\text{CBF Feedback}}+\underbrace{H_\pi^{-1}\delta\left(\hat{x}(t)\right)\alpha_\pi y_{\mathrm{lim}}^{\mathrm{min/max}}}_{\text{CBF Command}}
\end{aligned}
\tag{24}
$$

where the $i^{\mathrm{th}}$ component of the constant CBF command vector $y_{\mathrm{lim}}^{\mathrm{min/max}} \in R^m$ is defined as

$$
\left(y_{\mathrm{lim}}^{\mathrm{min/max}}\right)_i = \begin{cases} \left(y_{\mathrm{lim}}^{\mathrm{min}}\right)_i, & \text{if } \left(-H_x\hat{x}(t)-H_\pi u_{\mathrm{bl}}(t)+\alpha_\pi\, y_{\mathrm{lim}}^{\mathrm{min}}\right)_i > 0 \\ \left(y_{\mathrm{lim}}^{\mathrm{max}}\right)_i, & \text{if } \left(H_x\hat{x}(t)+H_\pi u_{\mathrm{bl}}(t)-\alpha_\pi\, y_{\mathrm{lim}}^{\mathrm{max}}\right)_i > 0 \\ 0 &, \text{ otherwise} \end{cases}
\tag{25}
$$

It can thus be seen that (7) can be written as

$$
\pi(t) = -\underbrace{H_\pi^{-1}\delta\left(\hat{x}(t)\right)\left(H_x-H_uK\right)}_{K_{i,\mathrm{CBF}}}\hat{x}(t)+\underbrace{H_\pi^{-1}\delta\left(\hat{x}(t)\right)\alpha_\pi}_{F_i}\, y_{\mathrm{lim}}^{\mathrm{min/max}} = -K_{i,\mathrm{CBF}}\hat{x}(t)+F_i y_{\mathrm{lim}}^{\mathrm{min/max}}.
\tag{26}
$$

Expressing the control augmentation in the form (21) shows the continuity of the proposed control design. Further, the equivalent form in (26) highlights that the proposed control augmentation remains closed-loop even when constraints are active. Finally, it can be seen that "infeasibilities/constraint violations" occurring, e.g., due to noise or disturbances, are systematically handled by the proposed method. While optimization-based methods need to pay special attention to address infeasibilities, the proposed formulation uses a QP only for design and thus, (26) provides a valid solution at all times. This aspect is addressed in Section VI with a gust rejection trade study.

Table I summarizes the overall derived servo-control augmentation design for clarity.

| | |
|---|---|
| Open-loop LTI MIMO dynamics (1) | $\dot{x} = Ax + Bu$ |
| Measured output (1) | $y = Cx + Du$ |
| Limited output (1) | $y_{\mathrm{lim}} = C_{\mathrm{lim}}\, x$ |
| State estimation error (8) | $e_{\mathrm{est}} = x - \hat{x}$ |
| Observer dynamics (6) | $\dot{\hat{x}} = A\hat{x} + Bu + L\left(y-\hat{y}\right)$<br>$\hat{y} = C\hat{x} + Du$ |
| Control input with baseline control and augmentation (3) | $u = u_{\mathrm{bl}} + \pi = -K\hat{x} + \pi$ |
| Operational constraints (2) | $y_{\mathrm{lim}}^{\mathrm{min}} \le y_{\mathrm{lim}} \le y_{\mathrm{lim}}^{\mathrm{max}} \Leftrightarrow h(x) = \begin{bmatrix} h_{\mathrm{min}}(x) \\ h_{\mathrm{max}}(x) \end{bmatrix} = \begin{bmatrix} y_{\mathrm{lim}}^{\mathrm{min}} - C_{\mathrm{lim}}x \\ C_{\mathrm{lim}}x - y_{\mathrm{lim}}^{\mathrm{max}} \end{bmatrix} \le 0$ |
| Modified operational constraints using state estimate (15) | $\begin{bmatrix} H_{\mathrm{min}}(\hat{x},u_{\mathrm{bl}},\pi) \\ H_{\mathrm{max}}(\hat{x},u_{\mathrm{bl}},\pi) \end{bmatrix} = \begin{bmatrix} -H_\pi \\ H_\pi \end{bmatrix}\pi + \begin{bmatrix} \Delta H_{\mathrm{min}}(\hat{x},u_{\mathrm{bl}}) \\ \Delta H_{\mathrm{max}}(\hat{x},u_{\mathrm{bl}}) \end{bmatrix} \le 0$<br>$\Delta H_{\mathrm{min}}(\hat{x},u_{\mathrm{bl}}) = -H_x\hat{x} - H_\pi u_{\mathrm{bl}} + \alpha_\pi\, y_{\mathrm{lim}}^{\mathrm{min}}$<br>$\Delta H_{\mathrm{max}}(\hat{x},u_{\mathrm{bl}}) = H_x\hat{x} + H_\pi u_{\mathrm{bl}} - \alpha_\pi\, y_{\mathrm{lim}}^{\mathrm{max}}$ |
| Auxiliary matrices (18) | $H_x = \begin{bmatrix} \left(C_{\mathrm{lim}}\right)_1 \prod_{j=1}^{r_1}\left(A-\lambda_{1j}I_n\right) \\ \vdots \\ \left(C_{\mathrm{lim}}\right)_m \prod_{j=1}^{r_m}\left(A-\lambda_{mj}I_n\right) \end{bmatrix}, \quad H_\pi = \begin{bmatrix} \left(C_{\mathrm{lim}}\right)_1 A^{r_1-1} \\ \vdots \\ \left(C_{\mathrm{lim}}\right)_m A^{r_m-1} \end{bmatrix} B$<br>$\alpha_\pi = \mathrm{diag}\left(\begin{bmatrix} c_{10} & c_{20} & \dots & c_{m0} \end{bmatrix}\right),\ c_{i0} = \prod_{j=1}^{r_i}\left(-\lambda_{ij}\right) > 0 \quad \forall i = 1,\dots,m$ |
| Min-norm optimal control augmentation (21) | $\pi = H_\pi^{-1}\left(\max\left(0_{m\times 1}, \Delta H_{\mathrm{min}}(\hat{x},\, u_{\mathrm{bl}})\right) - \max\left(0_{m\times 1}, \Delta H_{\mathrm{max}}(\hat{x},\, u_{\mathrm{bl}})\right)\right)$ |

**Table I. Min-norm Optimal CBF-based Servo-Control Augmentation Design Summary.**

## IV. Theoretical Properties: Stability, Margins, and Performance

### A. Stability and Boundedness

Since the control augmentation solution provides a continuous, piecewise-linear solution, closed-loop stability and boundedness can be assessed as in (10). Akin to the modified constrained output $Y_{\text{lim}}(t) = H_x x(t) + H_\pi u(t)$ in (15), we can define the constrained limited output using the state estimate

$$\hat{Y}_{\text{lim}}(t) = H_x \hat{x}(t) + H_\pi u(t) \tag{27}$$

By construction, the control augmentation enforces

$$\alpha_\pi \, y_{\text{lim}}^{\min} \le \hat{Y}_{\text{lim}}(t) \le \alpha_\pi \, y_{\text{lim}}^{\max} \tag{28}$$

We use (27) to express the control input by means of the state estimate and the bounded signal (28) with

$$u(t) = H_\pi^{-1}\left(\hat{Y}_{\text{lim}}(t) - H_x \hat{x}(t)\right) \tag{29}$$

Next, inserting (29) into the system dynamics (1), with $\hat{x}(t) = x(t) - e_{\text{est}}(t)$,

$$\begin{aligned} \dot{x}(t) &= Ax(t) + Bu(t) = Ax(t) + BH_\pi^{-1}\left(\hat{Y}_{\text{lim}}(t) - H_x \hat{x}(t)\right) \\ &= Ax(t) + BH_\pi^{-1}\left(\hat{Y}_{\text{lim}}(t) - H_x\left(x(t) - e_{\text{est}}(t)\right)\right) \\ &= \left(A - BH_\pi^{-1}H_x\right)x(t) + BH_\pi^{-1}H_x e_{\text{est}}(t) + BH_\pi^{-1}\hat{Y}_{\text{lim}}(t) \end{aligned} \tag{30}$$

The closed-loop system dynamics for baseline control and control augmentation are driven by the estimation error and the bounded signal (27), with the error dynamics being independent of the control augmentation,

$$\begin{bmatrix} \dot{x}(t) \\ \dot{e}_{\text{est}}(t) \end{bmatrix} = \begin{bmatrix} A - BH_\pi^{-1}H_x & BH_\pi^{-1}H_x \\ 0 & A - LC \end{bmatrix} \begin{bmatrix} x(t) \\ e_{\text{est}}(t) \end{bmatrix} + \begin{bmatrix} BH_\pi^{-1}\hat{Y}_{\text{lim}}(t) \\ 0 \end{bmatrix} \tag{31}$$

***Remark* 2**. Since the eigenvalues of $A - BH_\pi^{-1}H_x$ do not depend on the baseline control design, the ability to apply the proposed augmentation technique can directly be checked using (31). Further, the tuning parameters of the control augmentation (the eigenvalues in (18)) need to be chosen such that $A - BH_\pi^{-1}H_x$.

***Remark* 3**. It is clear from (31) that boundedness can be achieved using the proposed control augmentation for any baseline controller as long as $A - BH_\pi^{-1}H_x$ is Hurwitz.

### B. Robustness and Margins

For any servo-control design framework, robustness margins need to be analyzed. Here, we show how to assess gain and phase margins at the control input breakpoint, where the baseline control $u_{\text{bl}}(t)$ and the augmentation signal $\pi(t)$ are added to form the total control input $u(t)$. Using the notation in (24), the total control input can be written in terms of the continuous switching logic

$$u(t) = -\left(\left(K + K_{i,\text{CBF}}\right)\hat{x}(t) - F_i y_{\text{lim}}^{\min/\max}\right) \tag{32}$$

and the closed-loop system yields

$$\dot{x}(t) = Ax + Bu = Ax - B\left(\left(K + K_{i,\text{CBF}}\right)\hat{x}(t) - F_i y_{\text{lim}}^{\min/\max}\right). \tag{33}$$

The system loop gain transfer function matrix $L_u(s)$ can be computed with the total servo-controller in the form of (24), while zeroing out command/constant terms $y_{\text{lim}}^{\min/\max}$,

$$u_{\text{out}}(t) = -\left(K + K_{i,\text{CBF}}\right)\hat{x}(t) \tag{34}$$

For the output feedback case studied in this paper, $\hat{x}(t)$ and $x(t)$ are given by

$$\begin{aligned} \hat{x}(t) &= \left(s I_n - \left(A - B\left(K + K_{i,\text{CBF}}\right) - LC\right)\right)^{-1} LCx(t) \\ x(t) &= \left(s I_n - A\right)^{-1} Bu_{\text{in}}(t) \end{aligned} \tag{35}$$

Thus, inserting (35) into (34) yields

$$u_{\text{out}}(t) = -\left(K + K_{i,\text{CBF}}\right)\left(s I_n - \left(A - B\left(K + K_{i,\text{CBF}}\right) - LC\right)\right)^{-1} LC\left(s I_n - A\right)^{-1} Bu_{\text{in}}(t). \tag{36}$$

Finally, we can state the loop gain

$$u_{\text{out}}(t) = -\underbrace{\left(K + K_{i,\text{CBF}}\right)\left(sI_n - \left(A - B\left(K + K_{i,\text{CBF}}\right) - LC\right)\right)^{-1} LC\left(sI_n - A\right)^{-1} B}_{\text{Loop Gain}: L_u(s;\delta)} u_{\text{in}}(t) = -L_u(s;\delta)u_{\text{in}}(t). \tag{37}$$

In this case, SISO and MIMO margins at the system input breakpoint are defined based on the resulting $(m \times m)-$ dimensional loop gain transfer function matrix, parameterized with the binary-valued matrix $\delta$,

$$\begin{gathered} L_u(s;\delta) = \left(K + K_{i,\text{CBF}}\right)\left(sI_n - \left(A - B\left(K + K_{i,\text{CBF}}\right) - LC\right)\right)^{-1} LC\left(sI_n - A\right)^{-1} B \\ K_{i,\text{CBF}} = H_\pi^{-1}\delta\left(\hat{x}(t)\right)\left(H_x - H_\pi K\right) \end{gathered} \tag{38}$$

Hence, the loop gain (38) can be used to compute SISO and MIMO gain and phase margins [2] for all possible combinations of the binary-valued diagonal elements of $\delta$. For this analysis, it is assumed that $\delta$ is a constant diagonal matrix, with binary values on its diagonal.

### C. Constraint Satisfaction and Performance

In order to show constraint satisfaction, we use exponential decay/stability arguments of the state estimation error dynamics and of the operational constraint, i.e., when the modified constraint $H_i$ is active, the operational constraint $h_i = 0$ is being approached with exponential decay.

We present a sufficient condition for the state – not only the state estimate – satisfying the operational boundaries. Consider constraint $i$ with relative degree $r_i$ and CBF parameters $\{\alpha_{i0}, \alpha_{i1}, \ldots, \alpha_{i(r_i-1)}\}$, where $\alpha_{ij} = -\lambda_{ij}$ defining the eigenvalues of the stable polynomial (14). Theorem 1 with proof in the appendix shows that if at least one CBF parameter $\alpha_{ij}$ is chosen to be smaller than the slowest dynamics of the observer,

$$\begin{aligned} \alpha_i^\star &< \left|\lambda_{\max}^{\text{Re}}\right| \\ \alpha_i^\star &= \min\{\alpha_{i0}, \alpha_{i1}, \ldots \alpha_{i(r_i-1)}\} \\ \lambda_{\max}^{\text{Re}} &= \max_j \text{Re}\left(\lambda_j\left(A - LC\right)\right) \end{aligned} \tag{39}$$

where $\lambda_{\max}^{\text{Re}}$ is the maximum (slowest) eigenvalue of the observer, then there exists a finite time $t_{h\le 0}$ such that forward invariance is guaranteed forward in time, $h\left(x(t)\right) \le 0$ for all $t \ge t_{h\le 0}$. Theorem 1 implies that if the operational constraint is satisfied at any time $t_0$ during operation, it is guaranteed to be satisfied for all $t \ge t_0 + t_{h\le 0}$, i.e., after an initial transient in the estimation error has diminished. Note that Theorem 1 does not imply that the operational constraint will be violated. Assuming the initial estimation error to be sufficiently small, we can show that $t_0 = t_{h\le 0}$.

***Theorem 1 (Constraint Satisfaction).*** Consider the CBF augmentation computed using the state estimate in (21). Let the CBF parameter be chosen as in (39). Then, for any initial condition $x(t=0)$ with $h\left(x(t=0)\right) \le 0$, there exists a finite time $t_{h\le 0}$ such that $h\left(x(t)\right) \le 0$ for all $t \ge t_{h\le 0}$.

Fig. 2 illustrates the choice of CBF eigenvalues compared to observer eigenvalues. It shows a double-integrator model using only position measurements with velocity being the limited output and acceleration as the control input. We design an observer as in (6) with $L = \begin{bmatrix}4.0404 & 3.1623\end{bmatrix}^T$ and observer eigenvalues $\lambda_L = \{-2.9788, -1.0616\}$. Here, the modified constraint is given by

$$\dot{h} + \alpha h = -\dot{\hat{v}} + \alpha\left(v_{\text{lim}}^{\min} - \hat{v}\right) \le 0 \tag{40}$$

The CBF eigenvalues need to be chosen such that the state estimation error dynamics decay faster than the boundary of the CBF is being approached. Then, the state estimation error is upper bounded by the distance of the closed-loop system trajectory from the operational boundary. Fig. 2 illustrates two different scenarios: One where the CBF eigenvalues are chosen too aggressively, and one where the observer eigenvalues are used to correctly design the CBF parameter.

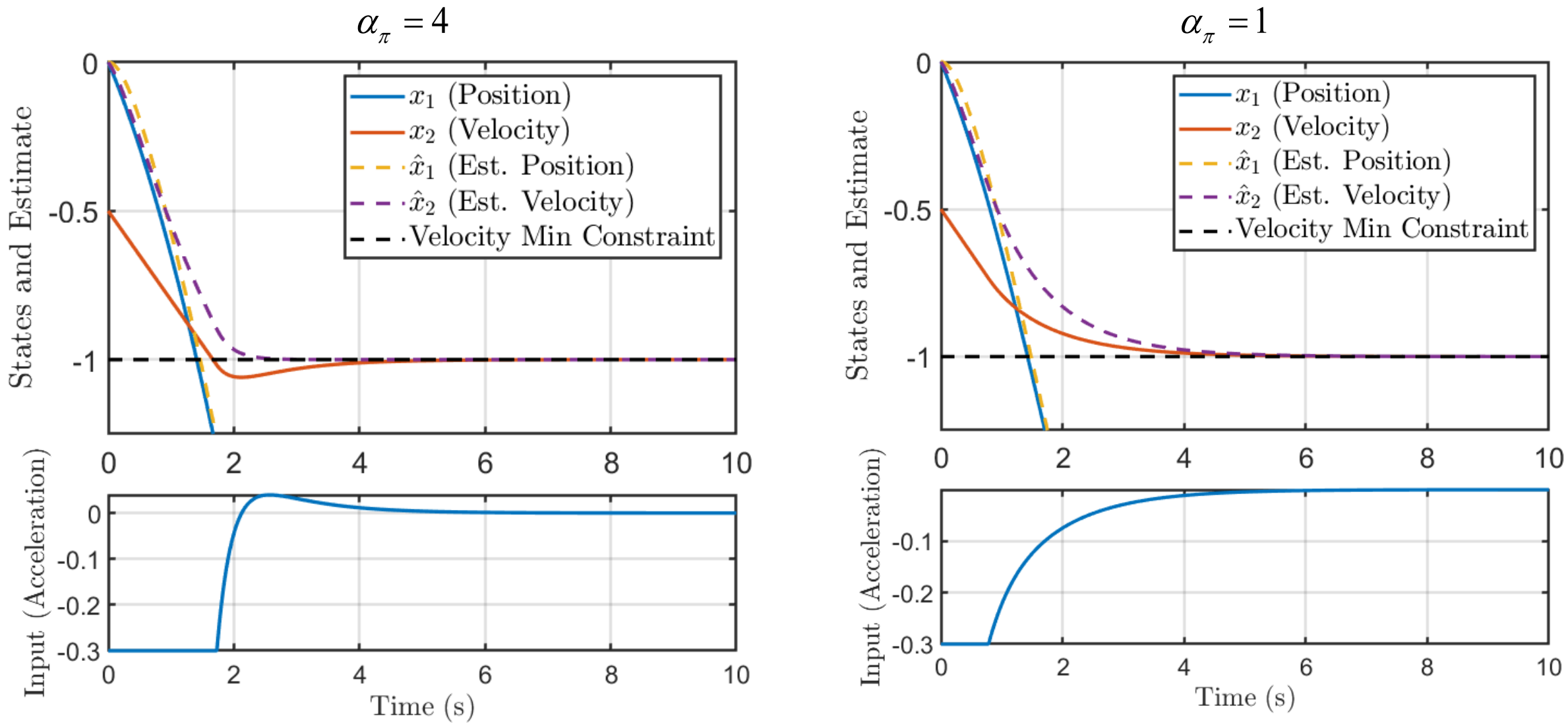

**Fig. 2. Illustration of impact of CBF parameter choice. Left: Incorrect choice of CBF parameters. Here, it is shown that the state estimation error converges slower than the boundary is being approached. In particular, the velocity (red) approaches the state estimate (purple dashed) slower than the state estimate (purple dashed) approaches the min. velocity constraint (dashed black). Consequently, the velocity (red) exceeds the operational boundary. Right: Correct choice of CBF parameters. Here, it is shown that the state estimation error converges faster than the boundary is being approached. In particular, the velocity (red) approaches the state estimate (purple dashed) faster than the state estimate (purple dashed) approaches the min. velocity constraint (dashed black).**

## V. Application to Proportional-Integral (PI) Controller

Consider the stabilizable MIMO LTI dynamical system,

$$\begin{aligned} \dot{x}_p(t) &= A_p x_p(t) + B_p u(t) \\ y(t) &= C_p x_p(t) + D_p u(t) \\ y_{\text{reg}}(t) &= C_{p\,\text{reg}} x_p(t) + D_{p\,\text{reg}} u(t) \\ z_{\text{lim}}(t) &= C_{p\,\text{lim}} x_p(t) \end{aligned} \tag{41}$$

where $x_p(t) \in R^{n_p}$ is the state vector, $u(t) \in R^m$ is the control inputs, $y(t) \in R^{n_y}$ is the vector of measured outputs, $y_{\text{reg}}(t) \in R^m$ is the vector of regulated outputs, and $z_{\text{lim}}(t) \in R^m$ is the vector of limited outputs. Here, we apply the generic design in Section III to a PI servo-controller with robust tracking of external commands $y_{\text{cmd}}$, while min/max operational constraints on the system control input $u(t)$ and on the selected limited outputs $z_{\text{lim}}(t)$ are satisfied for all times, component-wise,

$$\begin{aligned} u^{\min} &\le u(t) \le u^{\max} \\ z_{\text{lim}}^{\min} &\le z_{\text{lim}}(t) \le z_{\text{lim}}^{\max}. \end{aligned} \tag{42}$$

### A. Proportional-Integral Control Augmentation

Consider the control command,

$$u(t) = u_{\text{bl}}(t) + w(t) \tag{43}$$

with the proportional control augmentation signal $w(t)$. For robust tracking of external commands $y_{\text{cmd}}$ and operating in the presence of operational constraints, we use an integrated output tracking error $e_{yI}(t)$ augmented with the anti-windup (AW) control modification term $v(t) \in R^m$,

$$\begin{aligned} \dot{e}_{yI}(t) &= e_y(t) + v(t) \\ e_y(t) &= y_{\text{reg}}(t) - y_{\text{cmd}} \end{aligned} \tag{44}$$

with the output tracking error signal $e_y(t)$. Hence this paper designs a feedback control augmentation policy $w(t)$ and $v(t)$ such that the following properties are satisfied. The system's regulated output $y_{\text{reg}}(t)$ tracks external commands $y_{\text{cmd}}$ to the extent possible, while the control input $u(t)$ and the limited output $z_{\lim}(t)$ evolve within their operational constraints (42). Further, the state $x_p(t)$ and integrated output error $e_{yI}(t)$ remain bounded for all times.

### B. Augmented System Representation

For the PI controller design, the system is reformulated such that the generic control augmentation in Section III can be applied directly. First, augmenting the system dynamics in (41) with the error dynamics in (44) gives the $n=(n_p+m)-$ dimensional extended system

$$\underbrace{\begin{bmatrix}\dot{e}_{yI}(t)\\ \dot{x}_p(t)\end{bmatrix}}_{\dot{x}(t)}=\underbrace{\begin{bmatrix}0_{m\times m} & C_{p\,\text{reg}}\\ 0_{n\times m} & A_p\end{bmatrix}}_{A}\underbrace{\begin{bmatrix}e_{yI}(t)\\ x_p(t)\end{bmatrix}}_{x(t)}+\underbrace{\begin{bmatrix}I_m & D_{p\,\text{reg}}\\ 0_{n\times m} & B_p\end{bmatrix}}_{B}\underbrace{\left(\begin{bmatrix}-y_{\text{cmd}}\\ u_{\text{bl}}(t)\end{bmatrix}+\begin{bmatrix}v(t)\\ w(t)\end{bmatrix}\right)}_{\tilde{u}_{\text{bl}}(t)+\pi(t)}$$
$$\underbrace{\begin{bmatrix}e_{yI}(t)\\ y_p(t)\end{bmatrix}}_{y(t)}=\underbrace{\begin{bmatrix}I_m & 0_{m\times n}\\ 0_{n_y\times m} & C_p\end{bmatrix}}_{C}\underbrace{\begin{bmatrix}e_{yI}(t)\\ x_p(t)\end{bmatrix}}_{x(t)}+\underbrace{\begin{bmatrix}0_{m\times m}\\ D_p\end{bmatrix}}_{D}u(t). \tag{45}$$

Second, the limited output constraints and control input constraints are written in terms of the extended system,

$$y_{\lim}(t)=\begin{bmatrix}u_{\text{bl}}(t)\\ z_{\lim}(t)\end{bmatrix}=\underbrace{\begin{bmatrix}-K_I & -K_P\\ 0_{m\times m} & C_{p\lim}\end{bmatrix}}_{C_{\lim}}\begin{bmatrix}e_{yI}(t)\\ x_p(t)\end{bmatrix}=C_{\lim}\,x(t),\quad y^{\max}=\begin{bmatrix}u^{\max}\\ z_{\lim}^{\max}\end{bmatrix},\quad y^{\min}=\begin{bmatrix}u^{\min}\\ z_{\lim}^{\min}\end{bmatrix}. \tag{46}$$

Importantly, in [5], it is shown that limiting the baseline control policy $u_{\text{bl}}(t)$ in (46) implies that the total control command $u(t)$ evolves within the same bounds. Since (45) and (46) are now expressed in the form (1), the baseline controller in (5) and the control augmentation policy in (21) can be applied.

## VI. Flight Control Trade Studies

This section applies the proposed output feedback control method to a flight control example using PI control with command tracking. Consider the short-period dynamics (see [2], Section 14.7),

$$\begin{bmatrix}\dot{e}_{I,C^\star}(t)\\ \dot{\alpha}(t)\\ \dot{q}(t)\end{bmatrix}=\begin{bmatrix}0 & 6.78 & 1.29\\ 0 & -2.24 & 0.990\\ 0 & -4.47 & -0.902\end{bmatrix}\begin{bmatrix}e_{I,C^\star}(t)\\ \alpha(t)\\ q(t)\end{bmatrix}+\begin{bmatrix}0.876\\ -0.233\\ -4.59\end{bmatrix}\delta_{\text{ele}}(t)+\begin{bmatrix}-1\\ 0\\ 0\end{bmatrix}C^\star_{\text{cmd}}+\begin{bmatrix}0\\ -2.24\\ -4.47\end{bmatrix}\delta w_G(t)$$
$$\delta_{\text{ele}}(t)=\delta_{\text{ele,bl}}(t)+w(t) \tag{47}$$
$$C^\star_{\text{cmd}}=C^\star_{\text{cmd,bl}}-v(t)$$

with the angle of attack $\alpha(t)$, pitch rate $q(t)$, and integral regulated output error $e_{I,C^\star}(t)$. The regulated output is $C^\star(t)$, which is a linear combination of vertical load factor $N_z$ $(g)$ and pitch rate. The control input is $\delta_{\text{ele}}(t)$ and the command is $C^\star_{\text{cmd}}$. Additionally, we consider a disturbance/gust input $\delta w_G(t)$ (rad) impacting the closed-loop system as an angle of attack disturbance. The measured and limited output are

$$\text{Measured Output:}\quad y(t)=\begin{bmatrix}e_{I,C^\star}(t) & q(t)\end{bmatrix}^T=C\begin{bmatrix}e_{I,C^\star}(t) & \alpha(t) & q(t)\end{bmatrix}^T$$
$$\text{Limited Output:}\quad z_{\lim}(t)=\alpha(t)=C_{\lim}\begin{bmatrix}e_{I,C^\star}(t) & \alpha(t) & q(t)\end{bmatrix}^T. \tag{48}$$

The angle of attack and elevator input are bounded by

$$\alpha_{\lim}^{\min}=-5\deg\le\alpha(t)\le\alpha_{\lim}^{\max}=5\deg$$
$$\delta_{\text{ele}}^{\min}=-8\deg\le\delta_{\text{ele}}(t)\le\delta_{\text{ele}}^{\max}=8\deg \tag{49}$$

The controller design uses an LQR for both the feedback gain and Luenberger gain resulting in

$$K=\begin{bmatrix}-0.310 & -0.251 & -0.399\end{bmatrix}, L=\begin{bmatrix}3.16 & 9.29\\ 2.69 & 10.85\\ 0.947 & 5.92\end{bmatrix} \tag{50}$$

The resulting eigenvalues are

$$\begin{aligned}\text{eig}(A-BK)&=\{-1.52,-1.62+2.30\text{i},-1.62-2.30\text{i}\}\\ \text{eig}(A-LC)&=\{-3.28,-3.05+11.6\text{i},-3.05-11.6\text{i}\}\end{aligned} \tag{51}$$

As supported by our theoretical derivations, we choose the following CBF tuning parameter

$$\begin{aligned}\alpha_{\text{ele}}&=2\\ \alpha_{\text{AOA}}&=1.5.\end{aligned} \tag{52}$$

### A. Nominal Performance with Initial State Estimation Error

First, we show results without gust/disturbances with $\delta w_G(t)=0$. The initial state estimate and initial state are

$$\begin{aligned}&\hat{x}(t=0)=0\\ &x(t=0)=\begin{bmatrix}e_{I,C^*}(t=0) & \alpha(t=0) & q(t=0)\end{bmatrix}^T=\begin{bmatrix}0 & -4\deg & 0\end{bmatrix}^T\\ &e_{\text{est}}(t=0)=\begin{bmatrix}0 & -4\deg & 0\end{bmatrix}^T.\end{aligned} \tag{53}$$

#### 1. *Baseline Controller*

For comparison, Fig. 3 shows the performance of a baseline controller without control augmentation. The baseline controller does not keep the elevator commands within their limits, which is expected due to the absence of any modification. In order to track the command, the baseline controller produces an angle of attack that exceeds the operational boundaries. Note that the command is purposefully chosen to not be achievable with the given angle of attack and elevator limits in (49).

#### 2. *Controller with Proposed Augmentation*

In contrast, Fig. 4 shows the performance of the baseline controller with the proposed CBF augmentation. The proposed control augmentation prevents both input and output constraints from being violated. Further, the integral augmentation command modifies the command signal to a value that is physically achievable given the limits (49). This is not an ad-hoc modification. The integral augmentation has to change in order satisfy the operational constraints, which the proposed method does by construction. Overall, the control augmentation framework in this paper renders the system trajectories forward invariant with respect to its operational limits. The estimation error diminishes faster than the operational limit is being approached, which guarantees constraint satisfaction as proven in this paper.

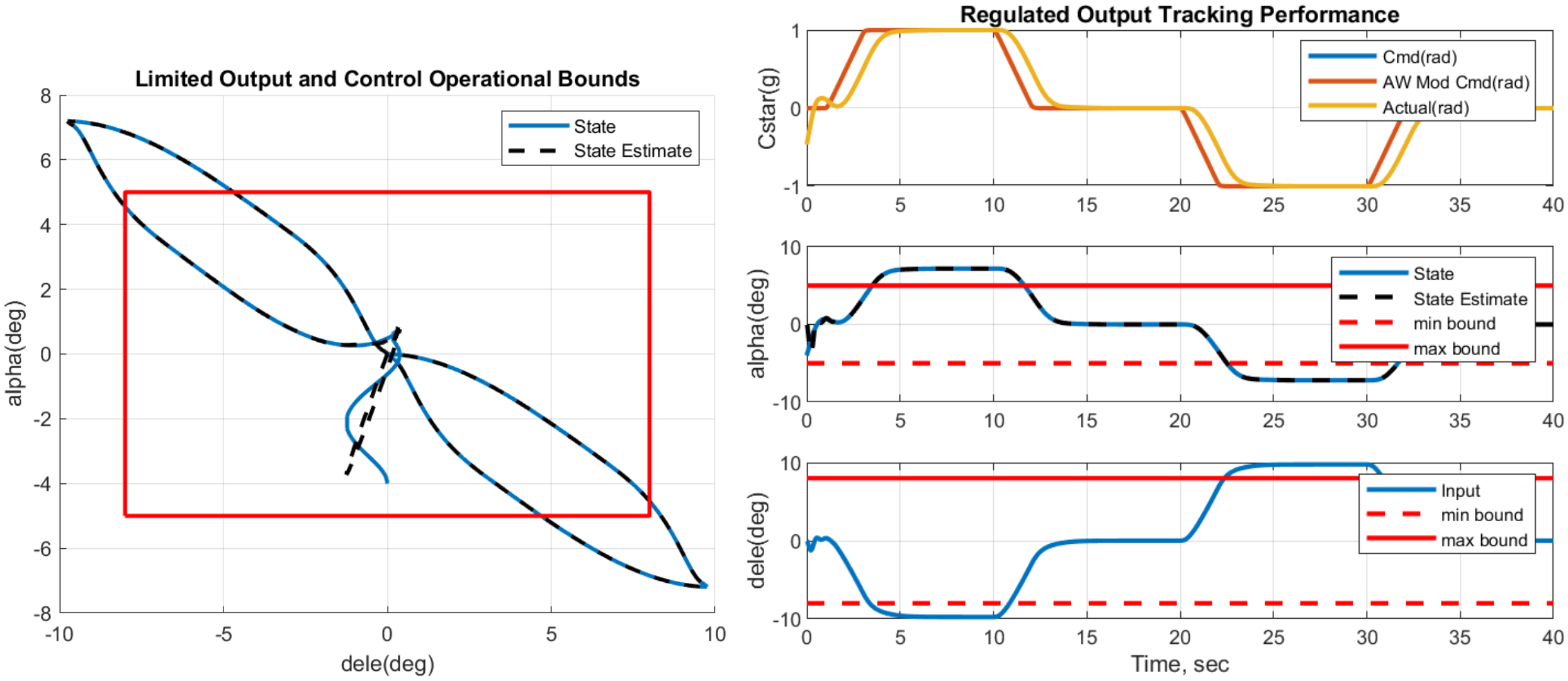

**Fig. 3. Command tracking with unconstrained baseline LQR PI controller. Left: Operational limits. Right: Aircraft tracking and constraint satisfaction performance. The baseline controller exceeds the operational control input and output limits (illustrated in red) in order to track the physically not attainable command.**

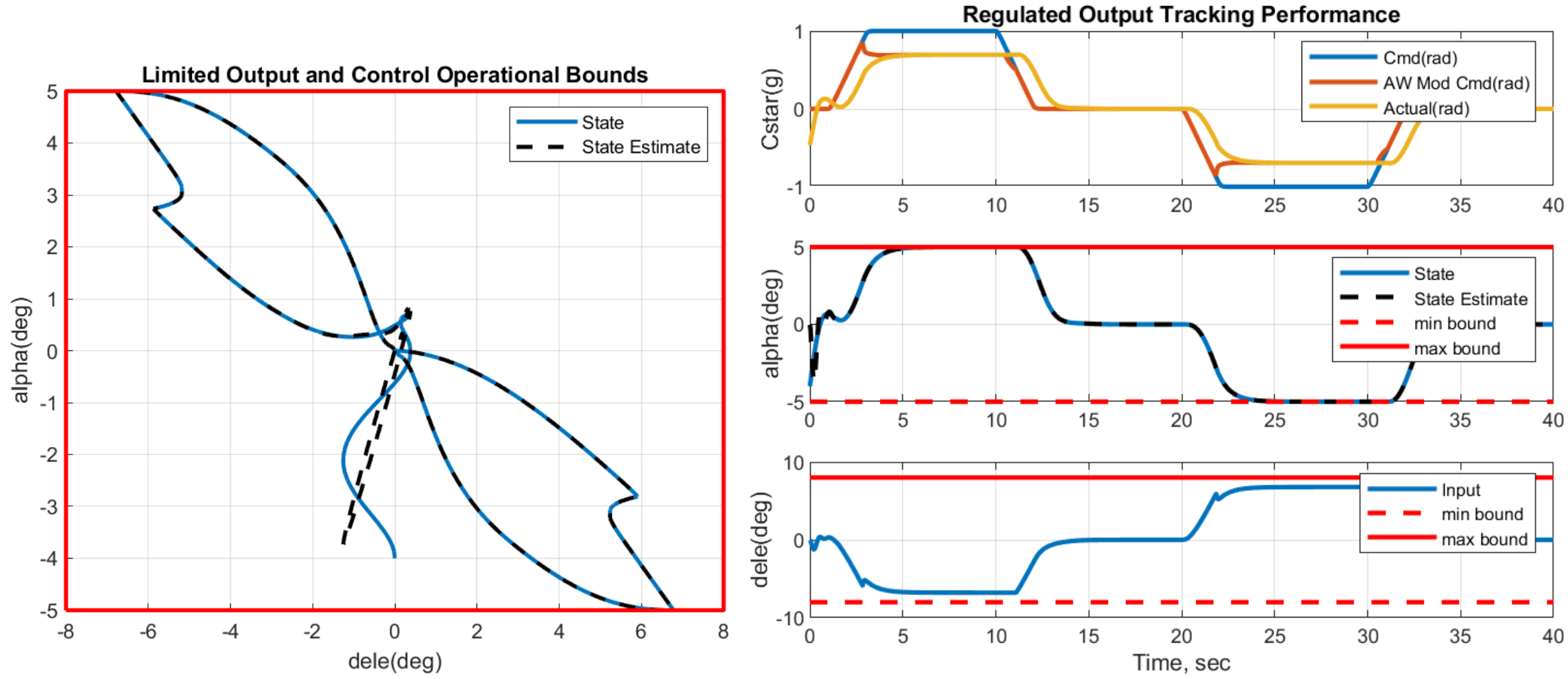

**Fig. 4. Command tracking with control augmentation. Left: Operational limits. Right: Aircraft tracking and constraint satisfaction performance. The control augmentation enforces both the control input and output limits. Due to these limits, the command can physically not be achieved. By construction, the proposed control augmentation modifies the command (from blue to red in top right plot) such that operational limits are respected.**

## B. Performance with Initial State Estimation Error and Gust

Here, we show with gust/disturbances modeled as in [2] (Fig. 14.8 Gust dynamic model, p. 608) and the same initial estimation error (53). Fig. 5 shows the resulting gust profile $\partial w_G$ .

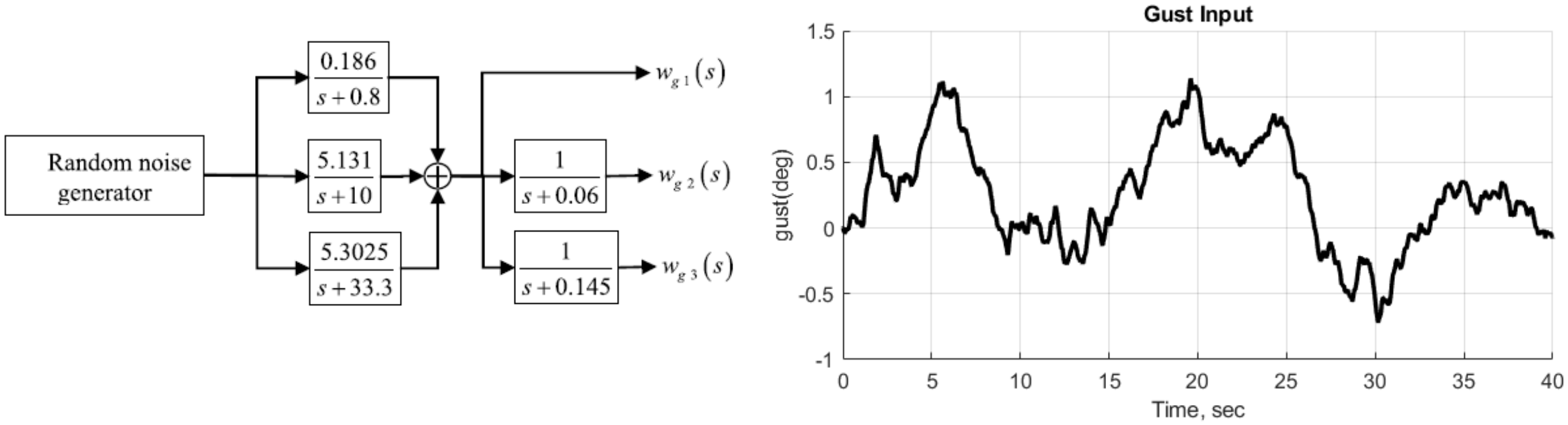

**Fig. 5. Left** [2]**: Gust generation, where we only use z-component $w_{g\,3}$ in this paper. Right: Resulting gust input.**

### *1. Baseline Controller*

Fig. 6 shows the performance of a baseline controller without control augmentation. Here, too, the baseline controller does not keep the elevator commands within their limits, which is expected due to the absence of any modification.

### *2. Controller with Proposed Augmentation*

Fig. 7 shows the performance of the baseline controller with the proposed CBF augmentation. Due to the gust/disturbance signal, the proposed control augmentation cannot guarantee constraint satisfaction for all times, which is no different from any other constrained control design. What is different in the proposed method, these temporary constraint violations do not cause infeasibilities since we use a QP only for design – not during execution. The control augmentation regulates (in closed loop) the angle of attack to the limits in (49). In practice, the limits enforced by the augmentation should be chosen sufficiently far away from the aircraft's physical limits in order to prevent physical limits from being exceeded.

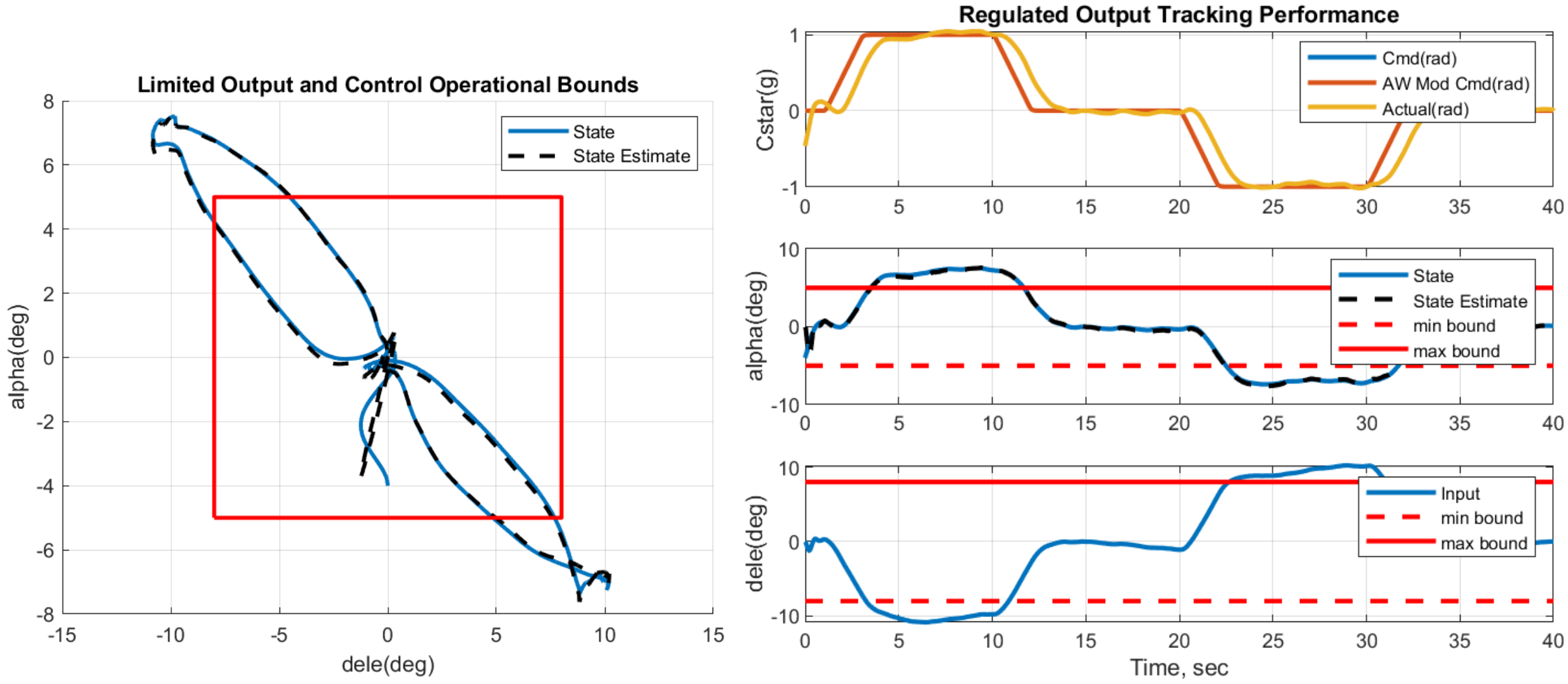

**Fig. 6. Command tracking with unconstrained baseline LQR PI controller with gust input. Left: Operational limits. Right: Aircraft tracking and constraint satisfaction performance. The baseline controller exceeds the operational control input and output limits (illustrated in red) in order to track the physically not attainable command. The baseline controller shows gust/disturbance rejection.**

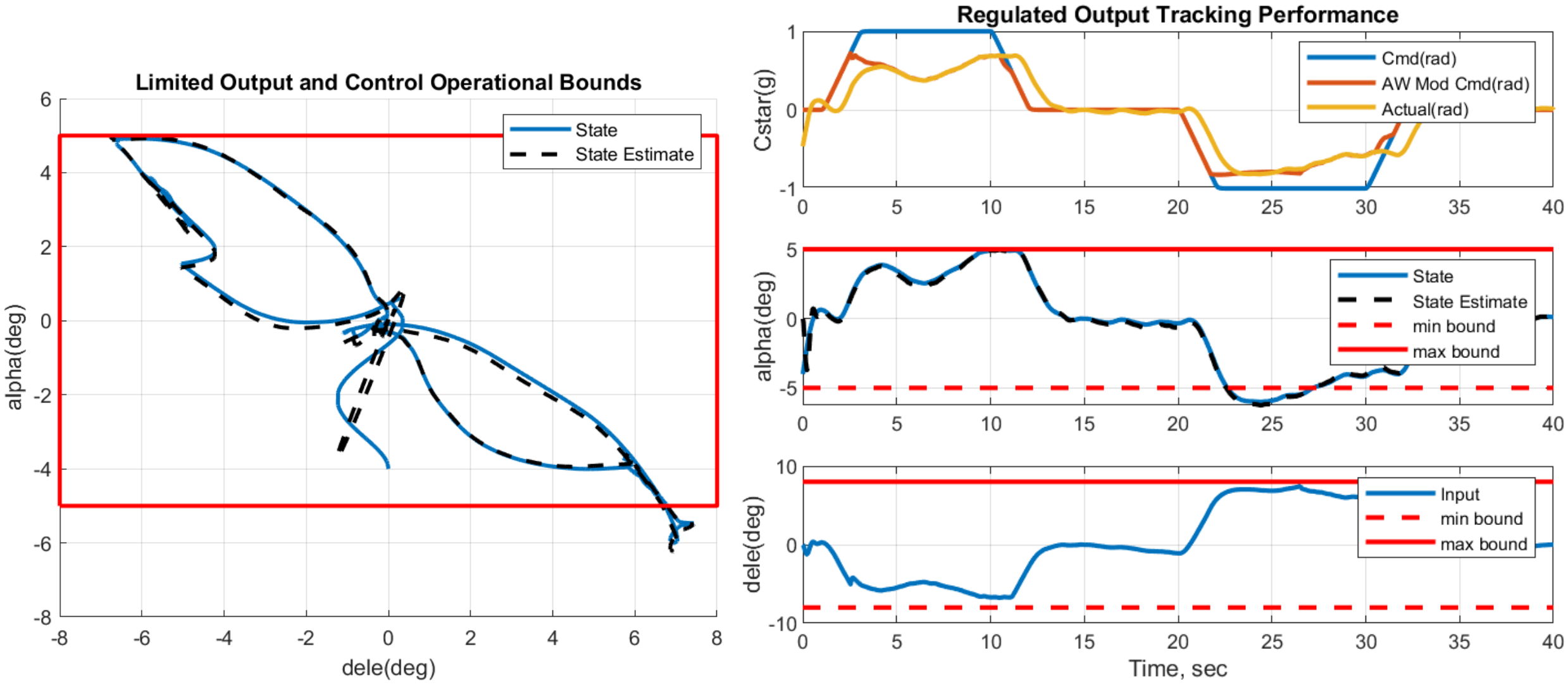

**Fig. 7. Command tracking with control augmentation and gust input. Left: Operational limits. Right: Aircraft tracking and constraint satisfaction performance. The control augmentation enforces both the control input and output limits. Due to these limits, the command can physically not be achieved. By construction, the proposed control augmentation modifies the command (from blue to red in top right plot). The control augmentation cannot guarantee constraint satisfaction due to the gust input.**

## C. Stability and Robustness Margins

One significant advantage of the proposed CBF augmentation design is its analyzability through linear systems theory. To illustrate this, Table II presents the MIMO margins computed at the input breakpoint. It shows the nominal gain and phase margins using the model in (47) as well as a scenario with second order (unmodeled) actuator dynamics,

$$\ddot{\delta}_{\text{ele}}(t) = -98\dot{\delta}_{\text{ele}}(t) + 4900\left(\delta_{\text{ele,cmd}} - \delta_{\text{ele}}(t)\right). \tag{54}$$

In the absence of active constraints, the proposed method is identical to the baseline control scenario, as indicated in the first row of Table II. However, when the elevator input reaches its limit, the proposed method with active aileron constraint remains closed-loop and exhibits comparable margins to the unconstrained case, while ensuring constraint satisfaction, as shown in the second row of Table II. The proposed control design also enforces limits on the angle of attack (AOA), which a baseline controller without modification cannot. Here, too, margins are comparable for the scenario with and without unmodeled actuator dynamics. This analysis underscores the robustness and effectiveness of the proposed CBF augmentation in managing control input constraints while preserving system stability. Fig. 8 illustrates the associated Bode diagram at the input breakpoint with actuator dynamics.

| Active Constraints | Baseline Controller | Baseline w/ Augmentation |
|---|---|---|
| one<br>None (w/ actuator dyn.) | GM = ∞, PM = 83.5deg<br>GM = 25.8dB, PM = 81.9deg | GM = ∞, PM = 83.5deg<br>GM = 25.8dB, PM = 81.9deg |
| Elevator<br>Elevator (w/ actuator dyn.) | N/A (constraint violated) | GM = ∞, PM = 102deg<br>GM = 30.4dB, PM = 101deg |
| AOA<br>AOA (w/ actuator dyn.) | N/A (constraints violated) | GM = ∞, PM = 129deg<br>GM = 48.0dB, PM = 113deg |
| AOA & Elevator<br>AOA & Elevator (w/ actuator dyn.) | N/A (constraints violated) | GM = ∞, PM = 131deg<br>GM = 48.8dB, PM = 116deg |

**Table II. Multi-Input Multi-Output (MIMO) margins for baseline controller without constraint enforcement and baseline controller with augmentation and constraint augmentation. Margins are computed for the nominal case as well as with unmodeled actuator dynamics.**

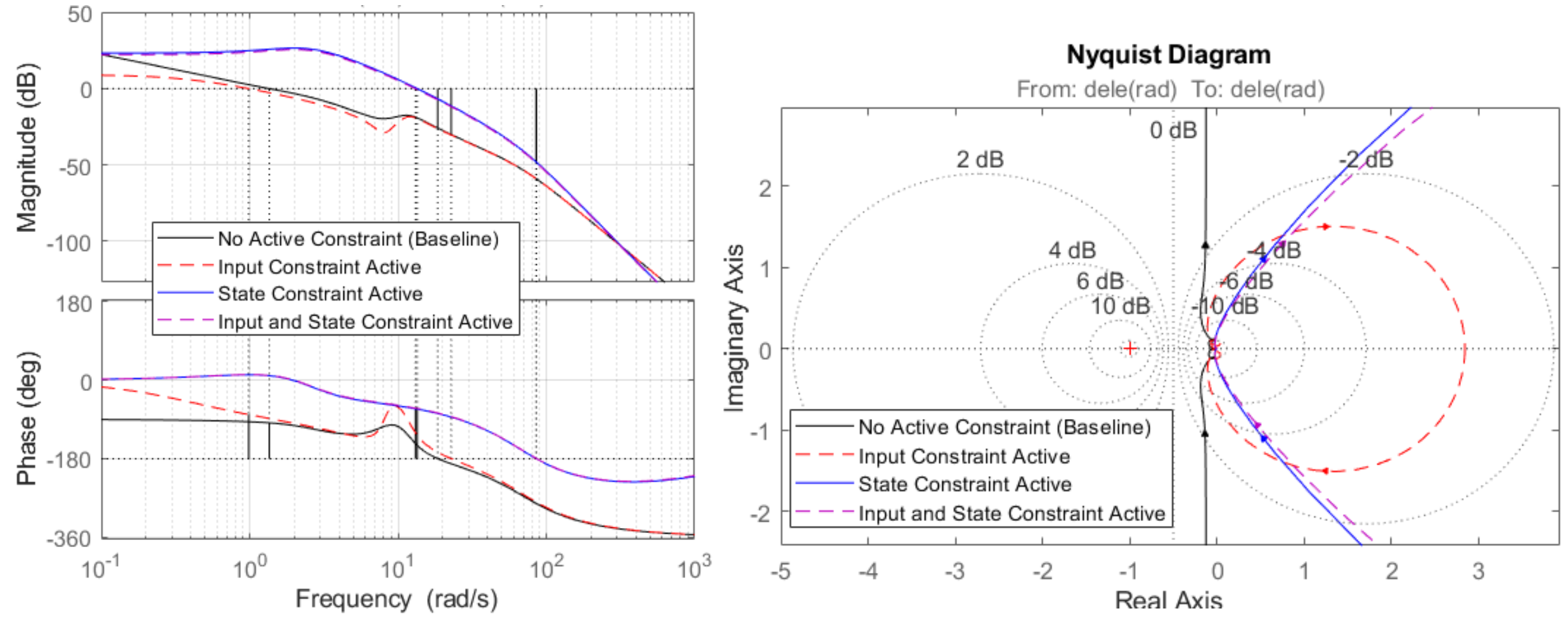

**Fig. 8. Bode diagram and Nyquist plot. Top left: Loop gain as function of frequency. Bottom left: Phase as function of frequency. Right: Nyquist plot.**

Finally, Fig. 9 presents gain and phase margins at the input breakpoint as a function of the augmentation parameters $\alpha_{\mathrm{AOA}}$ and $\alpha_{\mathrm{ele}}$. Such a plot and parameter study can be used for tuning. The parameters should not be chosen too small, which would yield conservative behavior and degraded margins. Larger parameters lead to a potentially too aggressive behavior, where margins also deteriorate. Notably, a sudden drop in both gain and phase margin can be observed at around $\alpha_{\mathrm{AOA}} = \alpha_{\mathrm{ele}} = 3$, which coincides with the real part of the slowest observer eigenvalues. This motivates the main result in this paper to choose the CBF parameters relative to the observer dynamics as in (39).

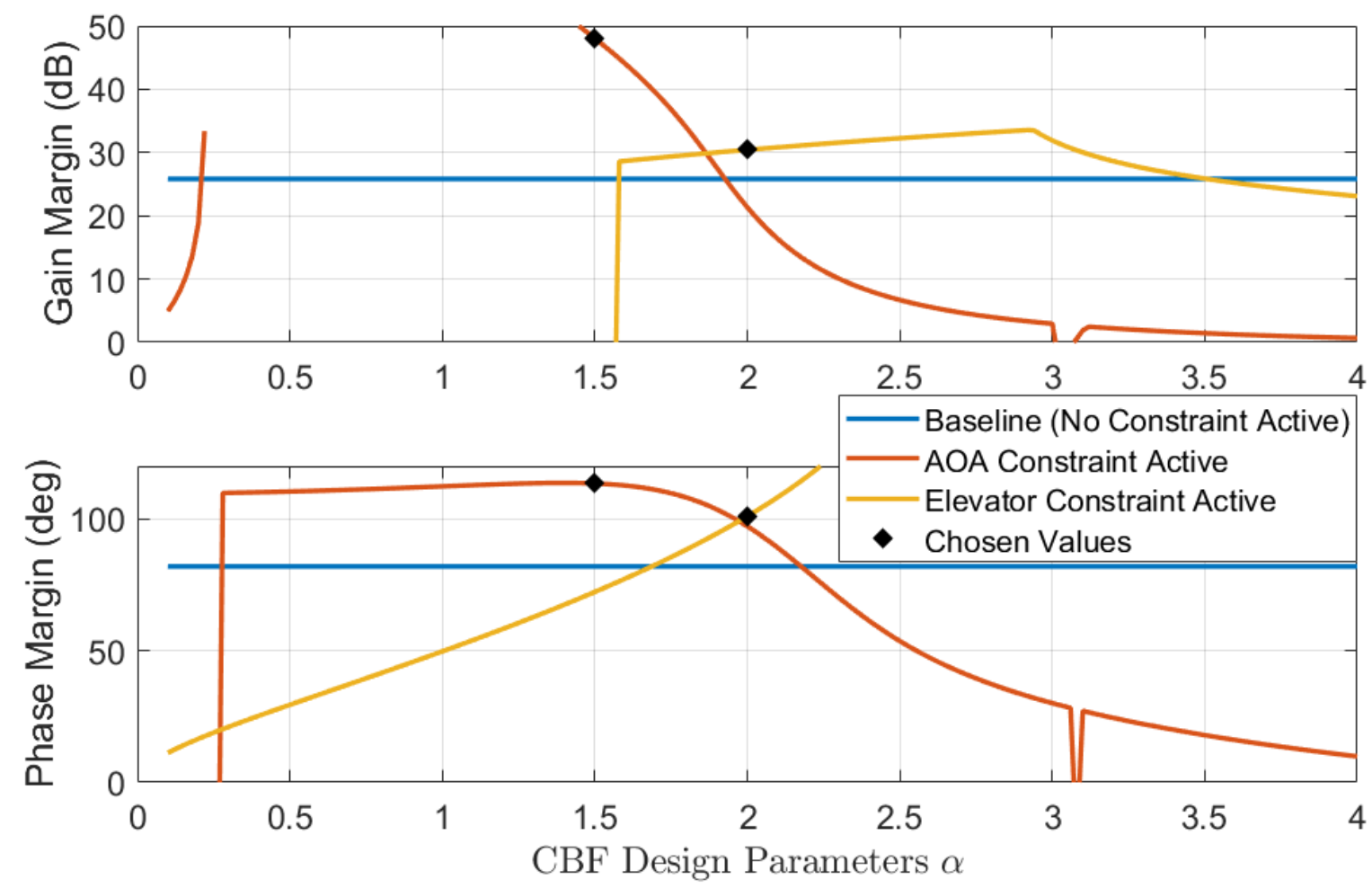

**Fig. 9. CBF design parameter study.**

## VII. Conclusions

In this paper, we have developed a systematic output feedback control design framework for robust linear servo-controllers that effectively enforce operational constraints. By integrating Control Barrier Function (CBF) theory with min-norm optimal control principles, the proposed method guarantees constraint satisfaction while preserving closed-loop analyzability with linear systems theory. Simulation results on a linear aircraft model demonstrate precise tracking of external commands under operational constraints using only output measurements. Our multi-input multi-output (MIMO) robustness margin analysis further reveals that the closed-loop system under output feedback maintains gain and phase margins close to those of the unconstrained baseline controller. This contrasts sharply with controllers relying on heuristic saturation handling, which suffer significant margin degradation and potential instability when constraints are active. Moreover, the paper also highlighted gust/disturbance rejection capabilities of the proposed method and presented robustness metrics to unmodeled actuator dynamics. Overall, these results highlight the advantages of our systematic output feedback control design, offering a rigorous and reliable framework to enhance robustness and safety in flight control systems subject to operational constraints.

## Appendix: Proof

***Proof of Theorem 1.*** By construction, for any relative degree $r_i$ of constraint $i$, the CBF augmentation enforces

$$\begin{bmatrix} -C_{\mathrm{lim},i}B \\ C_{\mathrm{lim},i}B \end{bmatrix}\pi(t)+\begin{bmatrix} -C_{\mathrm{lim},i}\left(A+\alpha_i^\star I_n\right)\hat{x}(t)-C_{\mathrm{lim},i}Bu_{\mathrm{bl}}(t)+\alpha_i^\star y_{\mathrm{lim},i}^{\min} \\ C_{\mathrm{lim},i}\left(A+\alpha_i^\star I_n\right)\hat{x}(t)+C_{\mathrm{lim},i}Bu_{\mathrm{bl}}(t)-\alpha_i^\star y_{\mathrm{lim},i}^{\max} \end{bmatrix}\le 0 \tag{55}$$

i.e., for relative degrees greater than one, $C_{\mathrm{lim},i}B=0$ but (55) still holds. This is true since the modified constraints in (15) are a more conservative inequality used for control augmentation design, i.e., any control augmentation signal designed using the modified constraints also satisfies (55). Replacing the state estimate yields

$$\begin{gathered}\begin{bmatrix} -C_{\mathrm{lim},i}B \\ C_{\mathrm{lim},i}B \end{bmatrix}\pi(t)+\begin{bmatrix} -C_{\mathrm{lim},i}\left(A+\alpha_i^\star I_n\right)\left(x(t)-e_{\mathrm{est}}(t)\right)-C_{\mathrm{lim},i}Bu_{\mathrm{bl}}(t)+\alpha_i^\star y_{\mathrm{lim},i}^{\min} \\ C_{\mathrm{lim},i}\left(A+\alpha_i^\star I_n\right)\left(x(t)-e_{\mathrm{est}}(t)\right)+C_{\mathrm{lim},i}Bu_{\mathrm{bl}}(t)-\alpha_i^\star y_{\mathrm{lim},i}^{\max} \end{bmatrix}\le 0 \\ \Updownarrow \\ \begin{bmatrix} -C_{\mathrm{lim},i}B \\ C_{\mathrm{lim},i}B \end{bmatrix}\pi(t)+\begin{bmatrix} -C_{\mathrm{lim},i}\left(A+\alpha_i^\star I_n\right)x(t)-C_{\mathrm{lim},i}Bu_{\mathrm{bl}}(t)+\alpha_i^\star y_{\mathrm{lim},i}^{\min} \\ C_{\mathrm{lim},i}\left(A+\alpha_i^\star I_n\right)x(t)+C_{\mathrm{lim},i}Bu_{\mathrm{bl}}(t)-\alpha_i^\star y_{\mathrm{lim},i}^{\max} \end{bmatrix}\le\begin{bmatrix} -C_{\mathrm{lim},i}\left(A+\alpha_i^\star I_n\right)e_{\mathrm{est}}(t) \\ C_{\mathrm{lim},i}\left(A+\alpha_i^\star I_n\right)e_{\mathrm{est}}(t) \end{bmatrix}\end{gathered} \tag{56}$$

It is easy to see that (56) can equivalently be written as

$$\begin{bmatrix} \left(s+\alpha_i^\star\right)h_{\min,i}\left(x(t)\right) \\ \left(s+\alpha_i^\star\right)h_{\max,i}\left(x(t)\right) \end{bmatrix}\le\begin{bmatrix} -C_{\mathrm{lim},i}\left(A+\alpha_i^\star I_n\right)e_{\mathrm{est}}(t) \\ C_{\mathrm{lim},i}\left(A+\alpha_i^\star I_n\right)e_{\mathrm{est}}(t) \end{bmatrix} \tag{57}$$

Next, consider the min constraint

$$\left(s+\alpha_i^\star\right)h_{\min,i}\left(x(t)\right)=\dot{h}_{\min,i}\left(x(t)\right)+\alpha_i^\star h_{\min,i}\left(x(t)\right)\le -C_{\mathrm{lim},i}\left(A+\alpha_i^\star I_n\right)e_{\mathrm{est}}(t) \tag{58}$$

We use the method of integrating factors and a series of conservative bounds. First, we multiply both sides of (58) by $e^{\alpha_i^\star t}$. Since $e^{\alpha_i^\star t}>0$, this does not change the direction of the inequality

$$\begin{gathered} e^{\alpha_i^\star t}\left(\dot{h}_{\min,i}\left(x(t)\right)+\alpha_i^\star h_{\min,i}\left(x(t)\right)\right)\le e^{\alpha_i^\star t}\left(-C_{\mathrm{lim},i}\left(A+\alpha_i^\star I_n\right)e_{\mathrm{est}}(t)\right) \\ \Updownarrow \\ e^{\alpha_i^\star t}\frac{\mathrm{d}h_{\min,i}\left(x(t)\right)}{\mathrm{d}t}+\frac{\mathrm{d}e^{\alpha_i^\star t}}{\mathrm{d}t}h_{\min,i}\left(x(t)\right)\le -e^{\alpha_i^\star t}C_{\mathrm{lim},i}\left(A+\alpha_i^\star I_n\right)e_{\mathrm{est}}(t) \end{gathered} \tag{59}$$

Second, we use the product rule to simply the expression on the left hand side of (59)

$$\frac{\mathrm{d}}{\mathrm{d}t}\left(e^{\alpha_i^\star t}h_{\min,i}\left(x(t)\right)\right)\le -e^{\alpha_i^\star t}C_{\mathrm{lim},i}\left(A+\alpha_i^\star I_n\right)e_{\mathrm{est}}(t) \tag{60}$$

Since (60) holds (in a differentiated sense), it also holds in an integrated sense,

$$e^{\alpha_i^\star t}h_{\min,i}\left(x(t)\right)\le h_{\min,i}\left(x(t=0)\right)+\int -C_{\mathrm{lim},i}\left(A+\alpha_i^\star I_n\right)e_{\mathrm{est}}(t)e^{\alpha_i^\star \tau}\mathrm{d}\tau \tag{61}$$

Next, it is clear that the state estimation error $e_{\text{est}}(t)$ decays exponentially and there exists a $k$ such that $C_{\lim,i}\left(A+\alpha_i^\star I_n\right)e_{\text{est}}(t)\le k\left\|e_{\text{est}}(t=0)\right\|e^{\lambda_{\max}t}$ for all $t\ge 0$ with the maximum (slowest) eigenvalue $\lambda_{\max}$. Note that $k$ can be explicitly stated using the system parameters. Thus,

$$e^{\alpha_i^\star t}h_{\min,i}\left(x(t)\right)\le h_{\min,i}\left(x(t=0)\right)+\int k\left\|e_{\text{est}}(t=0)\right\|e^{\lambda_{\max}\tau}e^{\alpha_i^\star\tau}\mathrm{d}\tau. \tag{62}$$

Integrating and dividing by $e^{\alpha_{0i}t}>0$ yields

$$\begin{aligned} h_{\min,i}\left(x(t)\right)&\le e^{-\alpha_i^\star t}h_{\min,i}\left(x(t=0)\right)+e^{-\alpha_i^\star t}\frac{k\left\|e_{\text{est}}(t=0)\right\|}{\left|\lambda_{\max}+\alpha_i^\star\right|}e^{\left(\lambda_{\max}+\alpha_i^\star\right)t}\\ h_{\min,i}\left(x(t)\right)&\le e^{-\alpha_i^\star t}h_{\min,i}\left(x(t=0)\right)+\frac{k\left\|e_{\text{est}}(t=0)\right\|}{\left|\lambda_{\max}+\alpha_i^\star\right|}e^{\lambda_{\max}t}\le 0.\end{aligned} \tag{63}$$

Finally, it is clear that a sufficient (conservative) condition for $h_{\min,i}\left(x(t)\right)\le 0$ is

$$\begin{aligned} h_{\min,i}\left(x(t)\right)\le e^{-\alpha_i^\star t}h_{\min,i}\left(x(t=0)\right)+\frac{k\left\|e_{\text{est}}(t=0)\right\|}{\left|\lambda_{\max}+\alpha_i^\star\right|}e^{\lambda_{\max}t}\le 0\\ \frac{k\left\|e_{\text{est}}(t=0)\right\|}{\left|\lambda_{\max}+\alpha_i^\star\right|}e^{\lambda_{\max}t}\le -e^{-\alpha_i^\star t}h_{\min,i}\left(x(t=0)\right).\end{aligned} \tag{64}$$

Rearranging the terms in (64) yields

$$e^{\left(\lambda_{\max}+\alpha_i^\star\right)t}\le -\frac{\left|\lambda_{\max}+\alpha_i^\star\right|}{k\left\|e_{\text{est}}(t=0)\right\|}h_{\min,i}\left(x(t=0)\right). \tag{65}$$

Clearly, for (65) to hold forward in time $t>t_{h\le 0}>0$, $\left(\lambda_{\max}+\alpha_{0i}\right)<0$ to keep the left hand side bounded. Finally, we can explicitly compute the time

$$t\ge t_{h\le 0}=\frac{1}{\left(\lambda_{\max}+\alpha_i^\star\right)}\ln\left(-\frac{\left|\lambda_{\max}+\alpha_i^\star\right|}{k\left\|e_{\text{est}}(t=0)\right\|}h_{\min,i}\left(x(t=0)\right)\right). \tag{66}$$

It can also be seen that for small initial state estimation errors $e_{\text{est}}(t=0)$, the logarithmic term can become positive,

$$\ln\left(\left\|\left(\lambda_{\max}+\alpha_{0i}\right)\right\|\frac{-h_{\min,i}\left(x(0)\right)}{\left\|C_{\lim,i}\left(A+\alpha_i^\star I_n\right)e(0)\right\|}\right)>0 \tag{67}$$

implying constraint satisfaction is guaranteed forward in time with $t_{h\le 0}=0$.